\documentclass[aip, jcp,  amsmath,amssymb]{revtex4-1}
\usepackage{bm}
\usepackage{float}
\usepackage{graphicx}
\usepackage{natbib}
\usepackage{color}

\begin{document}
\title{IETS and quantum interference: propensity rules in the presence of an interference feature}
\author{Jacob Lykkebo}
\affiliation{Nano-Science Center and Department of Chemistry, University of Copenhagen, Universitetsparken 5, 2100 Copenhagen \O, Denmark}
\author{Alessio Gagliardi}
\affiliation{Technische Universit\"at M\"unchen, Electrical Engineering and Information Tech., Arcisstr. 21, 80333 M\"unchen, Germany}
\author{Alessandro Pecchia}
\affiliation{Consiglio Nazionale delle Ricerche, ISMN, Via Salaria km 29.6, 00017 Monterotondo (Rome), Italy }
\author{Gemma C. Solomon}
\email{gsolomon@nano.ku.dk}
\affiliation{Nano-Science Center and Department of Chemistry, University of Copenhagen, Universitetsparken 5, 2100 Copenhagen \O, Denmark}
\date{\today}
%
%
\begin{abstract}
Destructive quantum interference in single molecule electronics is an intriguing phenomenon; however, distinguishing quantum interference effects from generically low transmission is not trivial. In this paper, we discuss how quantum interference effects in the transmission lead to either low current or a particular line shape in current-voltage curves, depending on the position of the interference feature. Secondly, we consider how inelastic electron tunneling spectroscopy can be used to probe the presence of an interference feature by identifying vibrational modes that are selectively suppressed when quantum interference effects dominate. That is, we expand the understanding of propensity rules in inelastic electron tunneling spectroscopy to molecules with destructive quantum interference.
\end{abstract}
\keywords{Inelastic electron tunneling spectroscopy, quantum interference, molecular electronics, propensity rules}
\maketitle
\section{Introduction}
Quantum interference in single molecule electronics is an intriguing phenomenon that is receiving considerable attention presently by experimentalists \cite{guedon2012,fracasso2011,arroyo2013, hong2011, aradhya2012, kaliginedi2012, ballmann2012} and from theoreticians \cite{solomon2008, solomon2008b, andrews2008, markussen2010, markussen2010b, markussen2011, markussen2014, hartle2011, ke2008, bergfield2009}. Destructive quantum interference can be understood as an anti-resonance in the Landauer transmission of the $\pi$-orbitals that leads to suppressed levels of current. However, it can be difficult to unambiguously distinguish between interference effects and a bad conductor with generally low levels of transmission. In a recent study, Guedon \textit{et al}. \cite{guedon2012} were able to observe the quantum interference directly in the conductance since the anti-resonance was positioned very close to the Fermi energy. However, for the same central molecular unit, but with a different linker group, the quantum interference in the conductance was not visible, and the only trace of the quantum interference was found in the generally lower levels of current. In another study, Aradhya \textit{et al}. were able to infer quantum interference effects in a meta substituted molecule, using force measurements even though the conductance of the molecule was below the detection level \cite{aradhya2012}. 
\\
In this paper, we discuss how line shapes of the transmission effect the resulting I(V) and G(V) (dI/dV) curves in the presence of an anti-resonance. We employ simple models to mimic different kinds of transmission line shapes and show how these affect the I(V) and G(V) curves. We conclude that the absence of the particular line shape does not necessarily exclude the possibly of quantum interference effects in a molecular junction. We then proceed to discuss another method to infer the existence of a quantum interference effect by using Inelastic Electron Tunneling Spectroscopy (IETS). We show and explain how particular vibrational modes are correlated with distance in energy between the anti-resonance and the Fermi energy, and how the normal propensity rules are affected by quantum interference effects. Thus, a chemical \cite{andrews2008} or an external gate can be used to probe the presence and position of the anti-resonance by IETS. The proximity of the anti-resonance to the Fermi energy is particularly interesting, since there seem to be a discrepancy between theoretical predictions and experimental findings \cite{guedon2012} and between different theoretical predictions on the same molecule (See for example \cite{guedon2012} and \cite{fracasso2011}).
\section{Methods}
The elastic and inelastic current  is calculated using the Meir-Wingreen equation. Inelastic contributions to the current are included within the Born approximation (BA) and solved up to third order, which was shown in a previous publication to be a good approximation to the full self-consistent Born for our system \cite{lykkebo2013}. The zeroth and first order terms in the Born equation are interpreted as the Landauer term, and additionally we have an inelastic term that involves absorption of a single quanta of energy by a vibrational mode. The current is given by
\begin{equation}
I = \int_{\mu_R}^{\mu_L} \mbox{Tr}[\mathbf{\Gamma}_L\mathbf{G}^r(E)\mathbf{\Gamma}_R\mathbf{G}^a(E)]dE + \int_{\mu_R+\omega_q}^{\mu_L}\mbox{Tr}[\mathbf{A}_L(E)\bm{\alpha}_{q}\mathbf{A}_R(E-\omega_{q})\bm{\alpha}_{q}]dE \mbox{  + higher order terms},
\label{current_eq}
\end{equation} 
where the first term is the elastic (Landauer like) term, and the second term is due to the (first order) inelastic current. $\mathbf{\Gamma}_{L/R}= -2$Im$\Sigma_{L/R}^{r}$ is the coupling to the left/right electrode in the wide-band limit, $\mathbf{G}^{r,a} = ((E-i\delta)-\mathbf{S}_{D}-\mathbf{H}_{D}-\mathbf{\Sigma}_L-\mathbf{\Sigma}_R-\mathbf{\Sigma}_{ph})^{-1}$, where $\delta$, $\mathbf{S}_D$, $\mathbf{H}_D$ is a positive infinitesimal,  the overlap and Hamiltonian matrix of the device region, respectively, and $\mathbf{\Sigma}_{L/R/ph}$ denote the self-energies due to the left and right leads and to the phonons, respectively. The spectral functions \cite{gagliardi2007} $\mathbf{A}_{L/R}$ are defined as $\mathbf{A}_{L/R}=\mathbf{G}^{a/r}\mathbf{\Gamma}_{L/R}\mathbf{G}^{r/a}$, and the eigenfunctions of $\mathbf{A}_{L/R}$ depict the coupling weighted density of states of the molecule \cite{gagliardi2007}. $\mathbf{\alpha}_{q}$ is the electron-phonon coupling matrix (el-p matrix) for mode q with phonon frequency $\omega_{q}$.  Diagonalizing the spectral functions by the unitary transformation $\bar{\mathbf{A}}_{L}(E)=\mathbf{C}^{\dagger}\mathbf{A}_{L}\mathbf{C}$, allows the transmission channels to be plotted. The transmission channels depend on energy, and take the form of the underlying molecular orbital at resonance. Off resonance, the elastic channels evolve with energy. 
\\
We note that there is a step discontinuity due to the approximation that $d^2I/dV^2 \approx dE/dV$. This implies that the peak heights depend on the energy grid used in the evaluation of equation \ref{current_eq}. The relative peak heights within a single calculation don't change, but in order to compare peak heights across different molecules it is essential to keep the grid spacing constant and this is done throughout this paper.
\\
\\
The inelastic (first-order) process can, by the second term in equation \ref{current_eq}, be interpreted a current entering the molecule from the left electrode and propagating through the molecule elastically ($\mathbf{A}_{L}$), followed by an inelastic scattering event due to the el-p matrix ($\mathbf{\alpha}_{q}$) and then finally elastically exiting the molecule into the right electrode, ($\mathbf{A}_{R}$). Furthermore, the inclusion of electron-phonon interactions also gives rise to elastic contributions, which is evident by the phonon self-energy term included in $\mathbf{G}^{r/a}$.
\section{Results}
\subsection{Interference effect on current-voltage and conductance voltage curves}
\begin{figure}[H]
\centering
\includegraphics[width=1.00\textwidth]{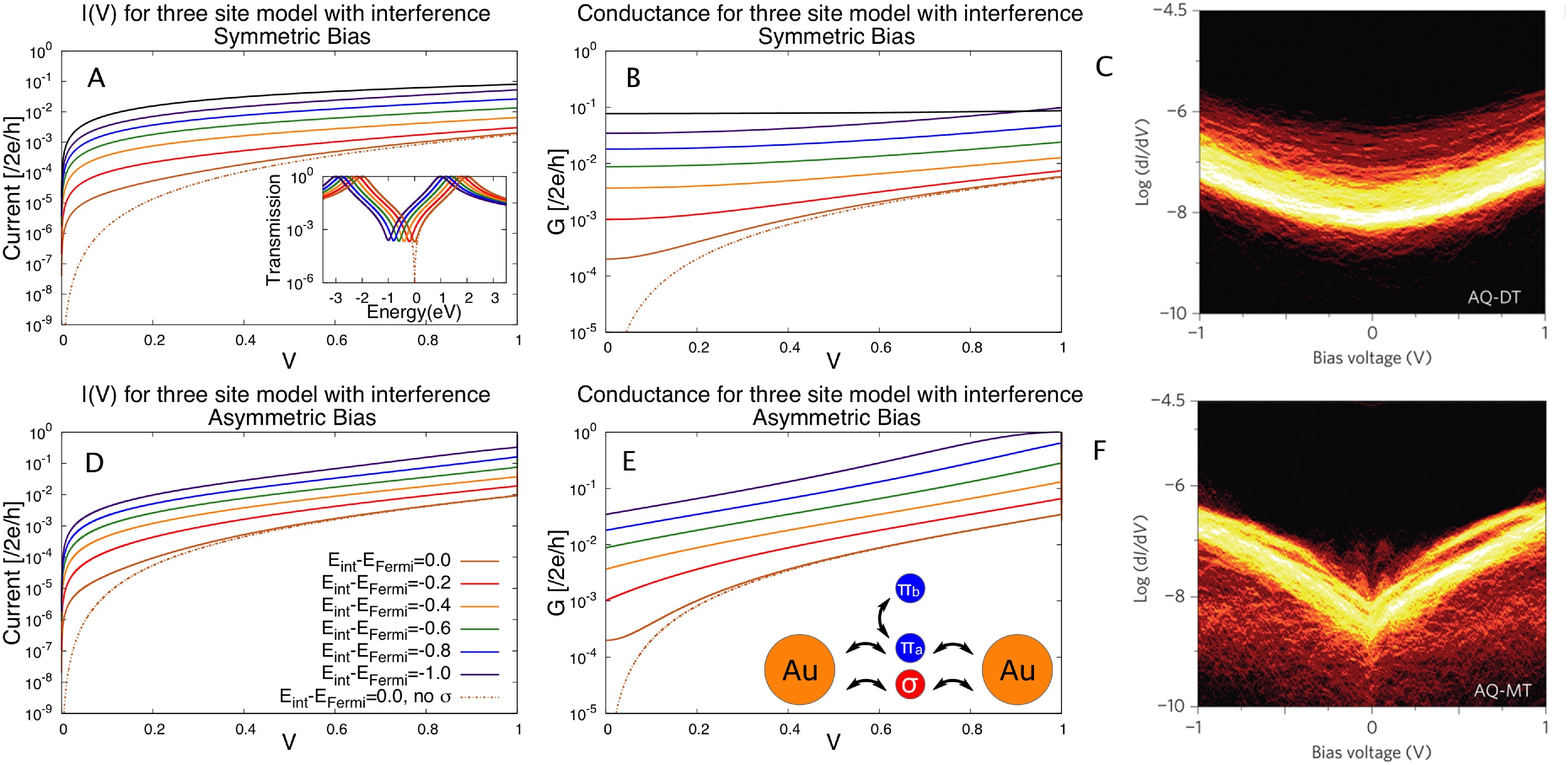}
\caption{Relationship between the line shape of the transmission function and the line shapes of the I(V) and on the conductance, for a symmetric and an asymmetric bias drop. $E_{int}-E_{Fermi}$ denotes the difference between the position of the anti-resonance and the Fermi energy, defined as the zero-point on the energy scale. For the dotted line the $\sigma$ site has been decoupled from the electrodes, reducing the system to a two site model. The parameters used for the site energies are: $\alpha_{\pi a}$=$\alpha_{\pi b}$=0, -0.1eV, -0.2eV, ..., -1.0eV, $\alpha_{\sigma}$=-10eV, $\alpha_{au}$=0. The parameters used for the couplings are: $\beta_{Au-\pi_a}$=-0.3eV, $\beta_{\pi a-\pi b}$=-2.0eV, $\beta_{Au-\sigma}$=-0.20eV. For the symmetric bias, the integration has been performed from -$V_{bias}/2$ to +$V_{bias}/2$ whereas for the asymmetric bias, the integration has been performed from 0 to $V_{bias}$. C and F show the log(G)(V) histograms for two molecules with the anti-resonance located at slightly different energies. The line shapes are observed to depend strongly on the location of the Fermi energy relative to the anti-resonance energy: Reprinted by permission from Macmillan Publishers Ltd: Nature Nanotechnology [\!\!\citenum{guedon2012}], copyright 2012.} 
\label{model_interference}
\end{figure}

We model a molecular junction with destructive interference effects using a simple tight-binding Hamiltonian with three sites in two orthogonal subsystems as illustrated in the inset to figure \ref{model_interference}E. These subsystems are designed to mimic the $\sigma$ and the $\pi$ systems. The position of the anti-resonance is controlled by the site energy of the $\pi$-orbitals, and by changing the site energy, the position of the anti-resonance moves. With this model we calculate the Landauer transmissions, (inset of figure \ref{model_interference}A), the I(V) curves (figure \ref{model_interference}A and D) and the conductances (\ref{model_interference}B and E). At the anti-resonance, the transmission through the $\pi$ subsystem goes to zero, so the value of the transmission minimum is determined by the $\sigma$ subsystem.  
\\
In reference [\!\!\citenum{guedon2012}], the conductances of two molecules with destructive interference effects was measured (Reprinted in figure \ref{model_interference}C and F), and one of the molecules exhibited a dip at low bias (Fig. \ref{model_interference}F). The qualitatively different line shapes were attributed to the fact that the anti-resonance of the transmission function was located at slightly different energies compared with the Fermi level. 
\\
When the anti-resonance in the transmission is exactly at the Fermi energy, the values of the I(V) and G(V) are very small at low bias, and the line shapes resemble the curves obtained in ref. [\!\!\citenum{guedon2012}] (See figure \ref{model_interference}F). Moving the anti-resonance slightly away from the Fermi energy significantly alters the I(V) and G(V) curves, and the dip at low bias rapidly disappears, which resembles the result in figure \ref{model_interference}C. The low levels of current and conductance persist, however. The special features in the I(V) and conductance curves are significantly reduced by moving the anti-resonance an energy corresponding to only 5\% of the HOMO-LUMO gap. 
\\
It should thus be noted that the presence of the particular line shape observed in ref. [\!\!\citenum{guedon2012}] probably stems from the fact that the anti-resonance is extremely close to the Fermi energy and the absence of the particular line shape does not necessarily exclude the possibility of destructive quantum interference. In some experimental setups, the voltage drop across the molecule is not symmetric due to pinning of the molecular levels to the substrate (\textit{e.g.} in STM measurements where the two electrodes differ tremendously in size). This does not change the quantitative effect on the line shapes of the I(V) curves or the conductance curves (Figure \ref{model_interference}C and D).
\\
\subsection{Using IETS to probe for interference effects} 
\begin{figure}[H]
\centering
\includegraphics[width=1.0\textwidth]{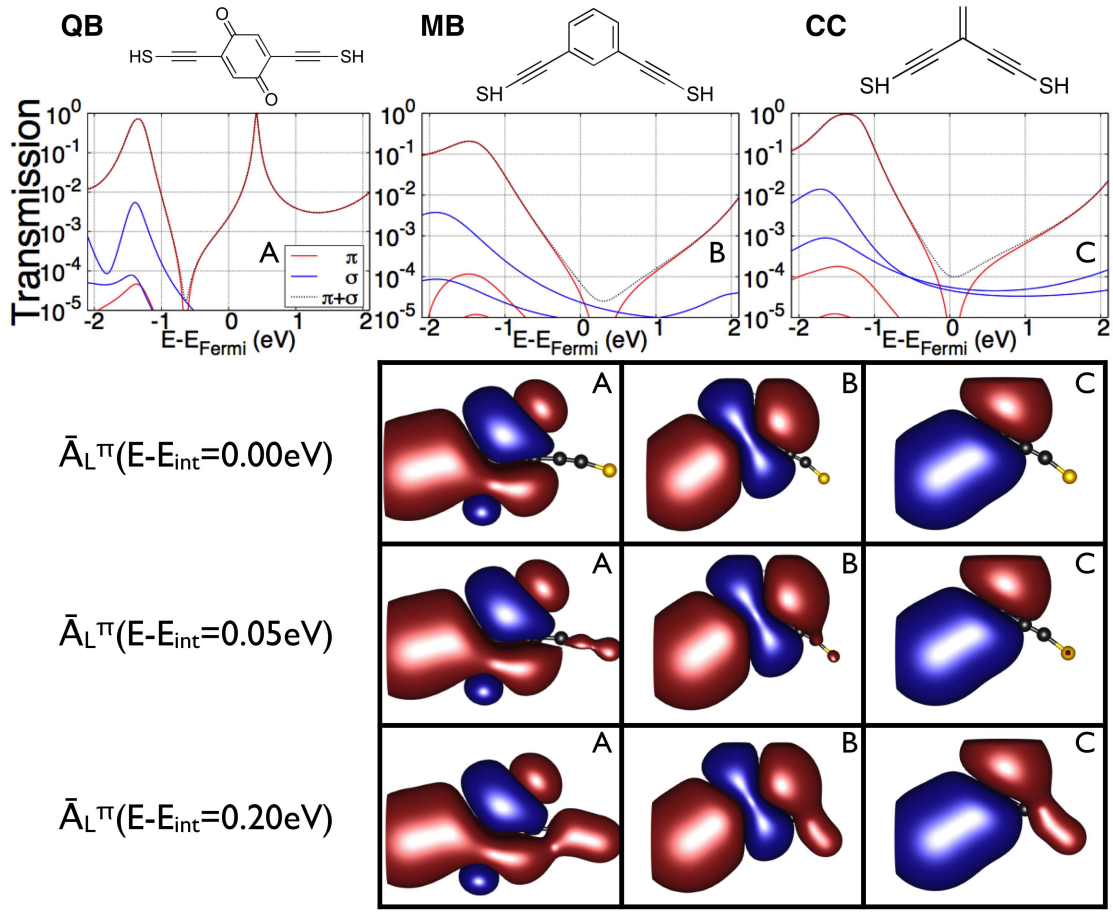}
\caption{Transmission and transmission channels ($\mathbf{\bar{A}}_{L}^{\pi}$) for \textbf{QB} (A), \textbf{MB} (B) and \textbf{CC} (C).The transmission channels are plotted at energies relative to the minimum of the $\pi$-
transmission.} 
\label{trans_chan}
\end{figure}

\begin{figure}[H]
\centering
\includegraphics[width=1.0\textwidth]{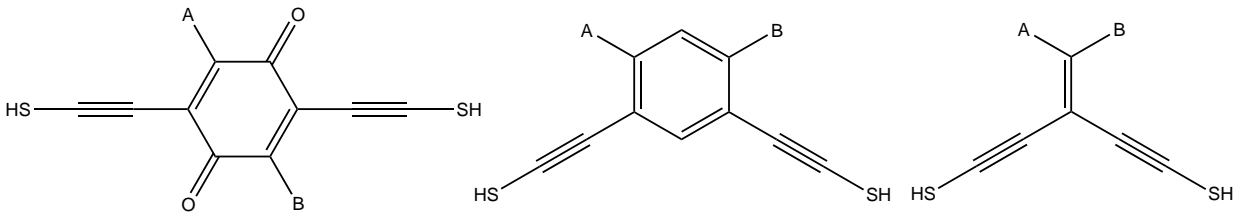}
\caption{The electron donating or -withdrawing groups are substituted with hydrogen in positions A or B, or both.}
\label{subs_strategy}
\end{figure}

\begin{figure}[H]
\centering
\includegraphics[width=1.0\textwidth]{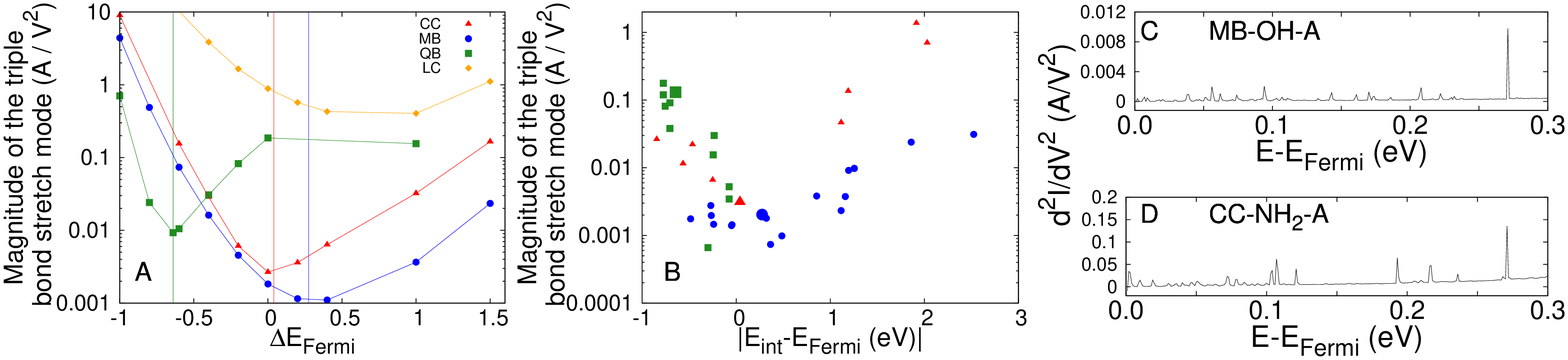}
\caption{The magnitude of the triple bond stretch when the position of the Fermi energy is moved manually away from its initial value of -4.8eV (A) and moved due to chemical substitutions (B). The vertical lines denote the position of the anti-resonance. C and D: IET spectra of $\textbf{MB}$ and $\textbf{CC}$ with substitutions. Triple bond stretch mode located at $\sim$0.27eV is the dominant mode. The position of the substituent for the molecules is A (see figure \ref{subs_strategy}).}
\label{fermi_and_chem_gate}
\end{figure}

\begin{figure}[H]
\centering
\includegraphics[width=1.0\textwidth]{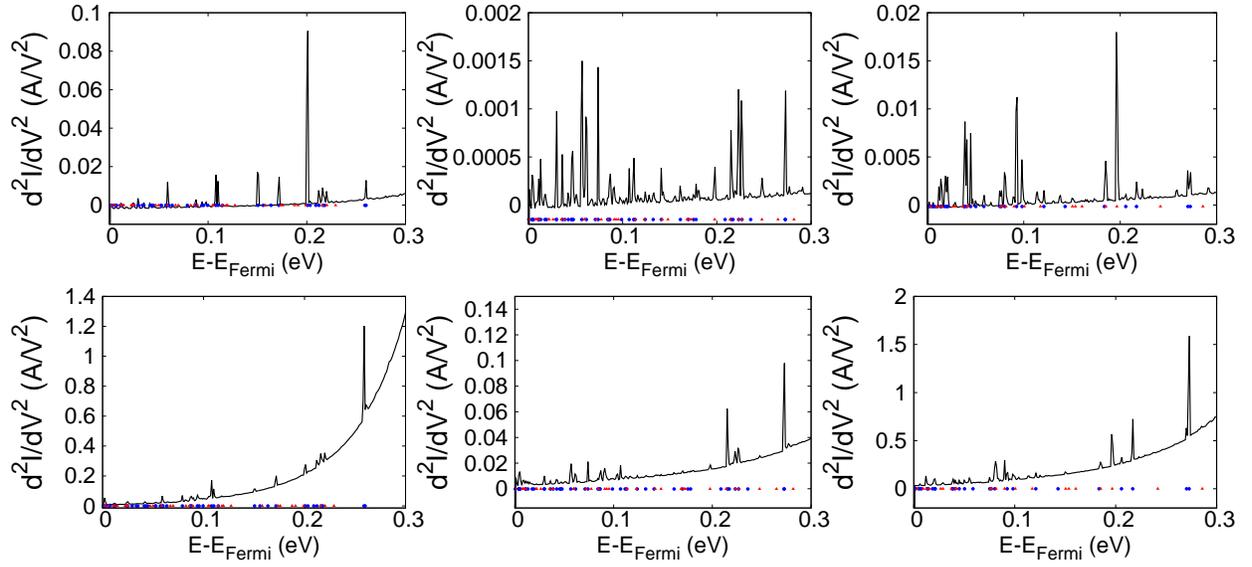}
\caption{IET spectra of $\textbf{QB}$, $\textbf{MB}$ and $\textbf{CC}$, respectively. The bottom row are calculated with the Fermi level 0.5eV below the energy of the interference feature. The top row are calculated with the Fermi level is equal to the energy of the interference feature. The triple bond stretch mode is located at $\sim$0.27eV.}
\label{IETS_at_Eint_and_away}
\end{figure}

If we cannot rely on the line shape of I(V) and G(V) , the question arises as to what other experimental signatures might be used to verify that low conductance is a result of destructive interference effects. In the second part of this paper, we illustrate how IETS can be used to probe the existence of an anti-resonance in the transmission. The potential for a peak in IETS arises when the excitation energy of a vibrational mode reaches the threshold $eV \ge \hbar \omega$. The peak height for a particular mode, however, depends strongly on the form of the electron-phonon matrix, and the form of the elastic channels as given by the second term in equation \ref{current_eq}. In the following, we show how destructive interference changes the IET spectrum.
\\
From the second term in equation \ref{current_eq}, it is clear that in order for an (First order) inelastic signal to exist, there needs to be an overlap between the channels $\mathbf{A}_{L}$, the el-p matrix and $\mathbf{A}_{R}$. For vibrational modes that are confined to a few atoms in the molecule, this implies that there needs to be overlap between the channels on these atoms. Otherwise, the inelastic current is always zero. Whether the peak from a particular vibrational mode is small or large is a more difficult question, since there are no hard selection rules to rely on\cite{gagliardi2007, troisi2006, troisi2006b}. 
\\
Many vibrational modes are confined to a specific part of the molecule, typically located on a few atoms. For localized vibrational modes, this implies that the channels must be overlapping on the few atoms where the vibration is located. For the molecules of this study, most of the vibrational modes are found to be localized stretch modes and benzene ring modes, while exceptions are mainly the low energy modes that correspond to translation and rotation of the entire molecule. This implies that for one such mode to be active in IETS, the transmission channels must have overlap on that part of the molecule. For ordinary linearly conjugated or saturated molecules, the channels are generally extended out all over the molecule so that no modes are strictly forbidden. However, for molecules exhibiting destructive interference effects, there can be a very small overlap between the left and the right channels \cite{lykkebo2013}, which can be shown directly by visualizing the elastic transmission channels. 
\\
\\
In this paper, three different molecules are employed in order to illustrate how quantum interference effects affect IETS in general way. The molecules are shown in figure \ref{trans_chan}, and are all characterized by an anti-resonance in the transmission near the Fermi energy. The molecules are chemisorbed (terminal hydrogens removed) to the fcc hollow site of two flat gold 111 surfaces. The molecule-electrode system is Cs symmetric, which allows for the separation of the $\sigma$ and $\pi$ transmission channels. In this paper, we use the Density Functional Tight Binding Method (See \textit{Computational Details}) which is know to give reasonable geometries\cite{bohr1999} and vibrational frequencies\cite{frauenheim2002}. It is generally found that in predicting quantum transport properties of molecular junctions DFTB is affected by similar errors as DFT, therefore I-V curves are of comparable quality. We demonstrated in previous works that electron-phonon couplings are sufficiently accurate to reproduce reasonably well relative peak strengths of IETS \cite{solomon2006}. We expect that the predictions of this work are trustable, especially when comparing trends and behaviors of different molecules. In figure \ref{trans_chan}, the transmission channels, $\bar{A}_{L}^{\pi}$,\cite{gagliardi2007} at various energies are shown for the three different molecules that exhibit destructive quantum interference effects in their transmissions. It can be seen that there is an anti-resonance in the $\pi$ transmission for all the molecules. The underlying transmission through the $\sigma$ system is unaffected by this, so the total transmission is finite even though the $\pi$ transmission goes to zero. The position of the anti-resonance relative to the Fermi energy differs slightly for all three molecules. The transmission at the anti-resonance is composed of contributions from two $\sigma$ channels for  \textbf{MB} and \textbf{CC}. The visualizations of the $\pi$ system channels $\bar{A}_{L}^{\pi}$ at the energy of the anti-resonance reveal the peculiar effect of the quantum interference. The density of the channel is high on the left side of the molecules, and thus illustrates that the left electrode is strongly coupled to the molecule. However, the channel density is very small on the right side of the molecule, and the current is therefore suppressed. When plotting the channels at energies slightly away from the anti-resonance, channel density builds up on the right side of the molecule, reflecting that the coupling from the left electrode begins to extend out across the whole molecule, and thereby also reflects an increase in current. Generally,  $\bar{A}^{\pi}_L(E')$ is identical to $\bar{A}^{\pi}_R(E')$, only mirrored so that there is overlap between the two channels only in the central region of the molecules. \textit{I.e.} there is no overlap between the linking groups of these molecules at the interference feature. The $\bar{A}^{\sigma}_L$ channel extends out across the whole molecule and is relatively constant in this range of energy. It is only at the anti-resonance that the $\sigma$ transmission is significant compared with the $\pi$ system.
The energy dependence in the form of the transmission results in a clear connection between the energetic distance between the interference feature and the Fermi energy, and the magnitude of the linking group mode in molecules with destructive quantum interference. We now focus our attention on the triple bond stretch mode. Using the triple bond stretch mode is a convenient choice, since many of the molecules studied in experiments are designed with triple bonds in the linker groups on either side of the central part of the molecule. This mode has been shown to be dominant in IETS for a linearly conjugated molecule \cite{kushmerick2004}. The triple bonds are suitable since they are rigid and therefore (hopefully!) ensure that the current flows exclusively through the molecule, and not \textit{via} some unknown short circuit or other unwanted route. \cite{hong2011, fracasso2011, arroyo2013, andrews2008, markussen2011}. Furthermore, the excitation energy of the triple bond stretch is $\sim$0.26eV, which is energetically quite isolated from the other vibrational modes, making it relatively easy to identify this mode in IETS. In comparison, the neighboring modes are the C-H stretches at $\sim$0.40eV and the double bond stretches at $\sim$0.20eV \footnote{It should be noted that DFTB is known to overestimate vibrational frequencies, which can be improved with an appropriate scaling factor. However, this overestimation is systematic so that the qualitative picture should still apply.}. 
\\
Furthermore, the el-p matrix for the triple bond stretch mode is essentially confined to the two pairs of carbon atoms involved in the bond. The overlap between the $\pi$ channels and the el-p matrix at the energy of the anti-resonance is therefore essentially zero. 
\\
The magnitude of the triple bond stretch mode is shown in figure \ref{fermi_and_chem_gate}, where the magnitude of the stretch mode is plotted as a function of distance in energy between the anti-resonance and the Fermi energy, calculated with gDFTB. The same qualitative result is obtained in two different ways. In figure \ref{fermi_and_chem_gate}A, the Fermi energy is manually changed, which in a simple fashion mimics the effect of an external gate. In figure \ref{fermi_and_chem_gate}B, the position of the anti-resonance is shifted by perturbing the electronic structure of the molecules by substituting hydrogen atoms with electron donating or -withdrawing groups\cite{andrews2008,markussen2011}. Here, the larger symbol in each of the three colors refer to the unsubstituted molecules. Both the external gate \cite{song2009b,prins2011,capozzi2014} and the chemical gate \cite{venkataraman2007} have been realized experimentally. Here cyano, alcohol, amine and nitro groups are used to effectively increase and decrease the on-site orbital energy of the neighboring carbon atom, which in turn changes the relative position of the anti-resonance. The functional groups are substituted with hydrogen on the sites marked with A and B (or both), as indicated in figure \ref{subs_strategy}. The substituents are found to mainly change the position of the anti-resonance, while leaving the positions of the resonances relatively unchanged. For the case of \textbf{MB}, it is found that substitutions in the positions meta to the linking groups have a negligible effect on the position of the anti-resonance. The ortho position that lies between the linking groups has an extremely large effect, but can even shift the anti-resonance out of the HOMO-LUMO gap. These substitution positions are therefore excluded from the study.
In figure \ref{fermi_and_chem_gate}A, where the Fermi energy is manually varied around the standard value of -4.8eV, it is observed that the minima in the magnitude of the triple bond stretch mode coincides with the position of the anti-resonance. In order to contrast the behaviour of molecules with quantum interference, we now compare with a the linearly conjugated molecule without quantum interference, \textbf{LC}. \textbf{LC} is the linearly conjugated analog of \textbf{CC} where the central ethylene unit has been rotated 90 degrees so that the molecule is fully conjugated from the left to the right electrode. The transmission of \textbf{LC} is therefore dominated by the $\pi$ system, with the transmission through the $\sigma$ system several orders of magnitude lower. \textbf{LC} shows a much smaller variation, and the magnitude of the triple bond stretch mode simply scales with the $\pi$ transmission. Furthermore, the triple bond stretch mode is predicted to be one of the dominant modes for this molecule \cite{lykkebo2013}. Conversely, in molecules dominated by destructive interference, we see that the observed magnitude of the stretch mode decreases when the anti-resonance approaches the Fermi energy. As the anti-resonance is moved away from the Fermi energy, this overlap between the left and right channels gradually increases, which causes an increase in the peak height of the mode. The peak height scales with the magnitude of the $\pi$ channels, so the peak height increases significantly as a resonance is approached. Thus, a steadily increasing peak height for $\textbf{QB}$ is not observed, after the resonance at $E_F+\approx 0.5eV$. 
\\
The same qualitative trend is observed irrespective of whether a chemical gate is used, or the Fermi energy is moved manually. For the chemical gate, figure \ref{fermi_and_chem_gate}B, the correlation is less clear since the substituents do more than simply move the anti-resonance, they also alter the geometry and general chemical structure of the molecules. However, the trend is still clear and it is observed that the peak height can change over several orders of magnitude using this approach.
\\
\\
The overall effect on the IET spectrum is studied by considering two cases: When the Fermi energy is exactly on the anti-resonance and when the Fermi level is 0.5eV lower than the antiresonance. The spectra are shown in figure \ref{IETS_at_Eint_and_away}, for $\textbf{QB}$, $\textbf{MB}$ and $\textbf{CC}$, respectively. When the Fermi energy and the anti-resonance coincide, all the modes from the linking groups are strongly suppressed due to the very small overlap between the left and right channels. On the other hand, when the Fermi level is moved away from the anti-resonance, the spectra change. It can be seen that the triple bond stretch mode changes from a small and insignificant mode, to the dominant feature for all three molecules. A careful analysis of the dominant modes in IET spectra when the Fermi energy is degenerate with the anti-resonance reveals a general trend: the symmetric (in-plane) modes are suppressed, while the asymmetric modes and overtones dominant the spectrum. The effect is the clearest for $\textbf{CC}$ and $\textbf{MB}$, where the double bond stretch mode of $\textbf{CC}$ (at $\sim$0.22eV) and the ring mode of $\textbf{MB}$ (at $\sim$0.22), are small compared with the overtones and the asymmetric modes \cite{lykkebo2013}. When the Fermi energy is moved 0.5eV down, these symmetric modes, along with the triple bond stretches, become the dominant modes. For $\textbf{QB}$ on the other hand, the symmetric C=O stretch at $\sim$0.20eV is the dominant mode with interference, but when the Fermi level has been moved away from the interference, this mode becomes almost completely silent. The slope of the baseline, is due to the shape of the transmission around the Fermi energy. Hence, is the baseline almost flat when the Fermi energy is set at the transmission minimum, whereas the slope increases when $E_{F}$ is halfway between a resonance and the anti-resonance.
The significant suppression of the triple bond stretch mode in the IET spectra, as a function of gate bias or a chemical gate, is a signature of interference effects in molecules. However, the minimum in the magnitude of the triple bond stretch mode vs. bias differs substantially for the three molecules. This stems from two different causes. 
\\
Firstly, the $\sigma$ system also contributes to the magnitude of the peak from the stretch mode, and the $\sigma$ system contributes to the total transmission with different magnitudes for the three different molecules. Specifically, the transmission from the $\sigma$ system of $\textbf{MB}$ is greater than that of $\textbf{CC}$, and this is the reason why the minimal value of the magnitude of the triple bond stretch mode vs. bias is greater for $\textbf{MB}$ compared with $\textbf{CC}$. 
\\
The second effect is that as the scattering process is inelastic, it is impossible to have both $\mathbf{A}_L(E)$ and $\mathbf{A}_R(E')$ located exactly at the anti-resonance, since the two channels are evaluated at energies separated by the phonon energy. Since the anti-resonance in the transmission of $\textbf{QB}$ is found to be quite narrow, compared with $\textbf{MB}$ and $\textbf{CC}$, the minimum of the magnitude of the triple bond stretch of $\textbf{QB}$ is not so low as it is for the other molecules with broader anti-resonance features. Still the magnitude of the mode is seen to decrease/increase 10 to 100 fold by using either chemical or an external gate to bring the anti-resonance closer to/further away from the Fermi energy. 
\\
Furthermore, the molecules used in this study are relatively small, so the contribution from the $\sigma$ system is generally larger than in related molecules used in experiments. This is because the current through the $\sigma$ system decreases faster with molecular elongation, compared with the current though the $\pi$ system \cite{cheng2011,moreno2013}.  Thus it is expected that the suppression of the mode here is probably underestimated, compared with the effect expected for the molecules that are employed in experiments. Indeed, elongating the $\textbf{MB}$ molecule by adding phenyl groups to the linkers, significantly decreases the minimum value of the magnitude of the triple bond stretch mode, because of the strong suppression of the $\sigma$ system (data not shown).
\subsection{Quantifying the role of the el-p matrix}
\begin{figure}[H]
\centering
\includegraphics[width=1.0\textwidth]{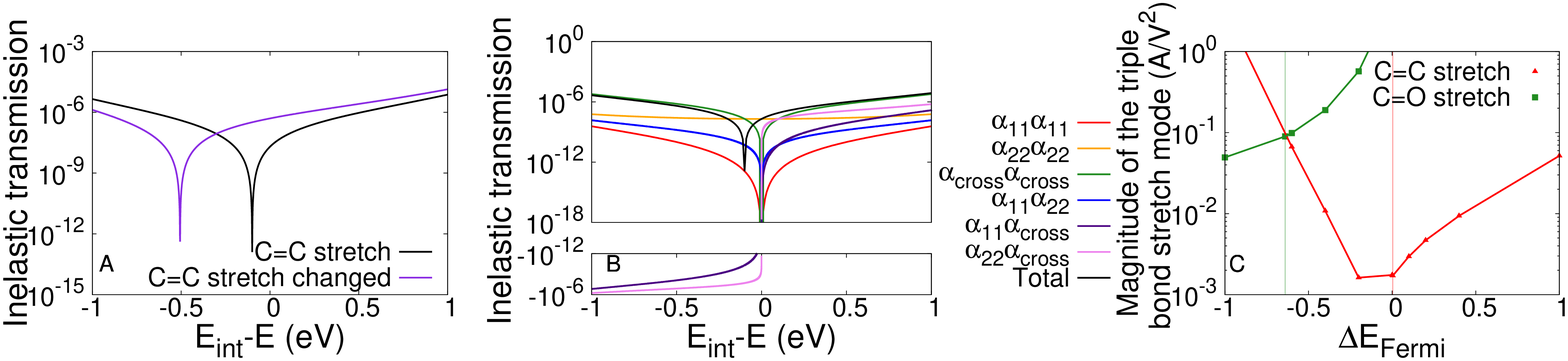}
\caption{\textbf{A}: Model calculation of the inelastic transmission calculated for a model of the CC double bond stretch mode using the standard (black) and an altered form (purple) of the el-p matrix  (see text) . The form of the el-p matrix determines the position of the anti-resonance in the inelastic transmission. \textbf{B}: Individual contributions to the inelastic transmission of the CC double bond stretch mode of \textbf{A}, where $\alpha_n\alpha_m = Tr[\mathbf{A}_L(E)\alpha_n\mathbf{A}_R(E)\alpha_m]$ (see text). A shift from negative to positive values are obtained for some of these contributions, which explains the local minima in the inelastic transmission. \textbf{C}: The magnitude of the carbon-carbon of \textbf{CC} and the carbon-oxygen stretch mode of \textbf{QB}, respectively. Vertical lines indicate the positions of the anti-resonances in \textbf{QB} (green) and \textbf{CC} (red), respectively.}
\label{central_modes}
\end{figure}

In the final part of the paper, we turn our attention to the vibrational modes of the central part of the molecules with quantum interference. At the anti-resonance, the modes in the linker groups are suppressed due to a lack of overlap between the two channels. However, many of these other in-plane modes are found to be suppressed due to the form of the el-p matrix, and not due to the lack of overlap of the channels. In particular, the symmetric modes in the benzene ring of \textbf{MB} and the double bond stretch mode of \textbf{CC} are only dominant modes away from the anti-resonance. On the other hand, the strong C=O stretch of \textbf{QB} is the dominant mode at the anti-resonance. From figure \ref{trans_chan}, it is seen that there is still channel overlap in the central part of the molecules, \textit{i.e.} the benzene ring of \textbf{QB} and \textbf{MB} and the ethylene group of \textbf{CC}, so it is clear that the fate of these modes is determined by the form of the el-p matrix.
\\
Let us now use the carbon-carbon double bond stretch mode of $\mathbf{CC}$ mapped onto the simple three site mode used previously to illustrate this. This mode is located at $\sim$0.21eV, and is observed to be suppressed at the Fermi energy (Figure \ref{IETS_at_Eint_and_away}C), but is one of the dominant modes away from the Fermi energy (Figure \ref{IETS_at_Eint_and_away}F) The form of the el-p matrix of this mode was found previously \cite{lykkebo2013}, and was found to reproduce quantitatively the results from a full DFTB calculation, using only the elements of the $\pi$-orbitals. The el-p matrix is 
\[\left(\begin{array}{cc} 
11 & 12 \\ 
21 & 22\end{array} \right)\  =  \left( \begin{array}{cc}
-0.001 & 0.01 \\
0.01 & -0.001 \\
\end{array} \right)\] 
where the (11) and (22) elements correspond to the sites $\pi$a and $\pi$b in the model in figure \ref{model_interference}, respectively, while the off diagonal elements corresponds to the coupling between them. Calculating the inelastic transmission for this mode (assuming a negligible phonon energy) yields the inelastic transmission shown in figure \ref{central_modes}A, which at a first glance look very similar to the results obtained above for the triple bond stretch mode, in the sense that the mode is suppressed at the anti-resonance energy. However, the reason why this mode is suppressed is due to the form of the el-p matrix, and by simply changing the  22 element of the el-p matrix from -0.001 to -0.005 moves the anti-resonance in the inelastic transmission away from the anti-resonance energy (Figure \ref{central_modes}A). An el-p matrix of this form suppresses the inelastic current somewhere away from the anti-resonance. Thus the symmetric mode will still be visible.
\\
The reason for the shift in anti-resonance in the inelastic transmission can be understood by a simple analysis of the different inelastic components through the molecule, which is obtained by splitting the el-p matrix up into a sum of sub-matrices,
\begin{equation}
\mathbf{\alpha}_q=\mathbf{\alpha}_{11} + \mathbf{\alpha}_{22} + \mathbf{\alpha}_{cross}
\end{equation}
where $\mathbf{\alpha}_{cross}$ consists of the off-diagonal terms. The (1st order) inelastic scattering term can now be split up into a sum of nine terms, which are the different components of the inelastic transmission. These are shown in figure \ref{central_modes}B. Some of the components are very small compared with the others (\textit{i.e.} irrelevant), some are positive and some are negative. For the parameters extracted for this el-p matrix these terms cancel very close to the position of the anti-resonance in the elastic transmission, which is why this mode is also suppressed in the IET spectra. For other symmetric vibrational modes in these types of systems, however, it is in no way guaranteed that the symmetric modes are suppressed. This is indeed what is obtained by studying the C=O stretch of $\mathbf{QB}$, as a function of the Fermi energy. This mode decreases in intensity as the Fermi level is decreased, but does not reach a minimum at the energy of the anti-resonance. This is clearly seen when plotting the peak height of the C=O stretch of \textbf{QB} and the C=C stretch of \textbf{CC} (Figure \ref{central_modes}C). 
The modes of the triple bonds are suppressed at the Fermi energy due to the non-overlapping channels, and this is irrespective of the form of the el-p matrix. Conversely, the symmetric carbon-oxygen stretch mode in $\mathbf{QB}$ is the dominant mode when the interference feature is found at the Fermi energy. However, the evolution of this mode when changing the Fermi energy differs substantially from that of the triple bond stretch mode, and illustrates how the dominant central modes are determined by the el-p matrix, whereas the dominant linker modes are determined mainly by the overlap of the transmission channels.
\subsection{Conclusion}
In conclusion, we show that quantum interference effects will manifest as low levels of current, unless the anti-resonance and the Fermi level are located close to each other. We furthermore discuss how inelastic electron tunneling spectroscopy is affected by quantum interference effects. Visually, quantum interference causes the elastic transmission channels to be abruptly truncated at the molecular motif that causes the interference effects. The transmission channels are thereby non-overlapping in large parts of the molecule, which causes the corresponding vibrational modes to be suppressed. Furthermore, many modes in the overlapping region are also suppressed, but this is shown to be caused by a cancelation of positively and negatively contributing terms. The results of this paper aid in the understanding of the interplay between current and molecular vibrations, and the interplay is shown to be fundamentally different depending on the position of the Fermi level and the anti-resonance. 
\\
The present understanding of the propensity rules that govern IETS for linearly conjugated and saturated molecules are shown to break down in the presence of an interference feature, leading to new propensity rules. In this paper we identify some of these, which could aid the study of the structure/properties relationship of single molecule junctions using IETS. 
\subsection{Computational Details}
All proper atomistic are calculated using the density functional tight binding approach. Geometries are obtained using the DFTB+ program \cite{porezag1995,seifert1996, elstner1998} and frequencies and transport calculations are obtained using gDFTB, density functional tight binding combined with the Green's function formalism \cite{pecchia2004, gagliardi2008, frauenheim2002}, while the electron-phonon coupling matrices are obtained using the method of reference \cite{pecchia2004b}. Starting geometries and parameters are obtained as in our previous work \cite{lykkebo2013}. 
\section{Acknowledgements}
The research leading to these results has received funding from the European Research Council under the European Union's Seventh Framework Program (FP7/2007-2013) / ERC Grant agreement n$^\circ$ 258806.
\bibstyle{achemso}
\end{document}